\documentclass{aastex61}

\newcommand\aastex{AAS\TeX}

\shorttitle{\aastex\ Further investigation of changes in cometary rotation}
\shortauthors{Mueller \& Samarasinha}
\begin{document}

\title{Further investigation of changes in cometary rotation}

\author[0000-0001-6194-3174]{Beatrice E.\ A.\ Mueller}
\affil{Planetary Science Institute \\
1700 East Ft.\ Lowell Rd., Suite 106\\
Tucson, AZ 85719, USA}

\author[0000-0001-8925-7010]{Nalin H. Samarasinha}
\affil{Planetary Science Institute \\
1700 East Ft.\ Lowell Rd., Suite 106\\
Tucson, AZ 85719, USA}

\begin{abstract}

\citet{SM13} related changes of cometary rotation to other physical parameters for four Jupiter family comets defining a parameter $X$, which is approximately constant within a factor of  two irrespective of the active fraction of a comet. Two additional comets are added to this sample in this paper and the claim of a nearly constant parameter $X$ for these six comets is confirmed, albeit with a larger scatter. Taking the geometric mean of $X$ for all the comets above excluding 2P/Encke (as $X$ for each comet was determined with respect to that of 2P/Encke), the expected changes in the rotation periods for a sample of 24 periodic comets are derived. We identify comets from this sample that are most likely to show observationally detectable changes in their rotation periods. Using this sample and including the six comets used to determine $X$, we find a correlation between the parameter $\zeta$ (i.e.\ the total water production per unit surface area per orbit approximated by that inside of 4\,au) and the perihelion distance $q$; specifically we derive $\zeta$\,$\propto$\,$q^{-0.8}$ and provide a theoretical basis for this in  Appendix A. This relationship between $\zeta$ and $q$ enables ready comparisons of activity due to insolation between comets. Additionally, a relationship between the nuclear radius $R$ and the rotation period $P$ is found. Specifically, we find that on average smaller nuclei have smaller rotation periods compared to the rotation periods of larger nuclei. This is consistent with expectations for rotational evolution and spin-up of comet nuclei, providing strong observational evidence for sublimation-driven rotational changes in comets.
\end{abstract}

\keywords{comets: general}

\section{Introduction} \label{sec:intro}

Theoretically, comets should change their rotational states due to torques acting on the nuclei when they are active. However,  in order to determine rotational changes, deriving accurate rotational states is observationally expensive and it would be beneficial to have a  predictor that could be used to assess which comets are likely to show observationally detectable changes in their rotation periods.

In a previous paper \citep[][hereafter SM13]{SM13}, we related changes in cometary rotation to activity for four Jupiter family comets, 2P/Encke, 9P/Tempel~1, 10P/Tempel~2, and 103P/Hart\-ley~2. There we defined a parameter $X$, and found that it was approximately constant  irrespective of the active fraction of a comet (see equation 12 in SM13). At that time, these were the only Jupiter family comets with reliably determined changes to their rotation periods. Currently, three additional Jupiter family comets have been confirmed with rotational changes, 19P/Borrelly \citep{MS15}, 41P/Tuttle-Giacobini-Kresak \citep[(41P/TGK)][]{Knight17,Schleicher17,Bode17,Bode18}, and 67P/Churyumov-Gerasimenko \citep[e.g.][]{Lowry12,Mottola14,esa17}. These objects other than comet 41P/TGK are used to reappraise  the parameter $X$. We  use the range of $X$ derived for these six comets and evaluate which other comets with observed rotation periods and known nuclear radii will be good candidates for observationally measurable period changes.
Furthermore, with a larger sample of comets with known rotation periods and nuclear radii, relationships between several orbital and physical parameters can be established.

Section~\ref{sec:paramX} discusses the reappraised parameter $X$ and the predicted changes in the rotation periods for 24 periodic comets, while section~\ref{sec:corr} shows relationships between various parameters. Section~\ref{sec:disc} discusses the results, and they are placed into a broader context, while section~\ref{sec:conc} summarizes the results. In the Appendix, we rationalize the relationship deduced in section~\ref{sec:corr} between two of the parameters. 

\section{Reappraisal of the parameter $X$ and predictions for rotational changes} \label{sec:paramX}

Table~\ref{tab:basecomets} lists Jupiter family comets  with confirmed period changes where $P$ is the rotation period and $\Delta P$ the change in the rotation period per orbit. A negative change in the period signifies a spin-up. The perihelion distances $q$, eccentricities $e$, and effective radii $R$  are also shown. Rotational values and effective radii for comets 2P/Encke, 9P/Tempel~1, 10P/Tempel~2, and 103P/Hartley~2 are from SM13, and references therein, values for comet 19P/Borrelly are from \citet{MS15}  and \citet{Lamy98}, and values for comet 67P/Churyumov-Gerasimenko are from \citet{Lowry12}, \citet{Mottola14}, \citet{esa17}, and \citet{Jorda16}. Comet 103P/Hartley~2 is in an excited spin state and the change in rotation period quoted in Table~\ref{tab:basecomets} corresponds to the dominant component period. We did not include comet 41P/TGK in Table~\ref{tab:basecomets} as  its spin state and spin evolution are not firmly established.  \citet{Bode18} reported  an increase  in rotation period  from 24\,hours to more than 42\,hours over two months, which is at least an order of magnitude higher than the period changes in the other comets.  However, this is in conflict with radar observations \citep{LH17}. Other possible explanations for this seemingly large change in the rotation period were also not explored in their paper, e.g.\ a non-principal axis spin state. In addition, the formalism developed for determining the parameter $X$ is not justifiable for comet 41P/TGK as we considered only changes in rotation periods which were much smaller than the rotation periods, i.e.\ $\Delta P$\,$\ll$\,$P$. 

We list the rotation periods in Table~\ref{tab:basecomets} to the nearest hour even though some of the rotation periods are determined more accurately, as we are taking the mean of the rotation period from before and after the change and more relevantly as we are only interested in determining $X$ to an accuracy smaller than its scatter which is about half an order of magnitude.

\begin{deluxetable*}{lccccc}[hbt]
\tablecaption{Perihelion distances $q$, eccentricities $e$, effective radii $R$, rotation periods $P$, and changes in rotation periods $\Delta P$ per orbit for the six Jupiter family comets with confirmed rotational changes \label{tab:basecomets}}
\tablecolumns{6}
\tablewidth{0pt}
\tablehead{
\colhead{\hspace{-20ex} comet name} &
\colhead{$q$} &
\colhead{$e$} &
\colhead{$R$} &
\colhead{$P$} &
\colhead{$\Delta P$} \vspace*{-1.75ex}  \\
\colhead{} &
\colhead{[au]} &
\colhead{} &
\colhead{[km]} &
\colhead{[hours]} &
\colhead{[min]} 
}
\startdata
2P/Encke & 0.34 & 0.85 & 2.4 & 11 & 4 \\
9P/Tempel 1 & 1.53  & 0.51 & 2.83 & 41 & -14 \\
10P/Tempel 2 & 1.42 & 0.54 & 6.0 &\ 9 & 0.27 \\
19P/Borrelly & 1.36 & 0.62 & 2.52 & 28 & $>$40 \\
67P/Churyumov-Gerasimenko & 1.24 & 0.64 & 1.65 & 12 & -21 \\
103P/Hartley 2 & 1.06 & 0.70 & 0.58 &18 & 150 \\
\enddata
\end{deluxetable*}

Table~\ref{tab:Xparam} lists the parameter $X$ as well as other related quantities with respect to comet 2P/Encke for the comets from Table~\ref{tab:basecomets} where $|\Delta P|$ is the absolute value of the change in rotation period per orbit, $R$ the  effective radius of the nucleus, $P$ the rotation period, and $\zeta$ the total water production per unit surface area per orbit based on the thermo-physical model of \citet{Pedro01}) approximated by that inside of 4\,au   (see SM13).  Note that in SM13, $\zeta$ was calculated to only 3\,au. $X$ is expressed as shown below (from equation 12 of SM13):
\begin{equation} \label{eq:X}
X = \frac{| \Delta P| R^2}{P^2 \zeta} 
\end{equation}
As can be seen from Table~\ref{tab:Xparam}, the parameter $X/X_{\rm E}$ is still constant within a factor of a few.  We conclude that {the parameter $X$ is still nearly constant}, albeit with a larger scatter, i.e.\ $X$ is nearly constant, particularly as  the active fractions  of these objects span 1.5 orders of magnitude  (SM13). We can now use this fact to estimate which other periodic comets with known rotation periods and radii are most likely to have an observable change in rotation period.

\begin{deluxetable*}{lccccc}[bt]
\tablecaption{The parameter $X$ and quantities relevant for its derivation for the comets from Table~\ref{tab:basecomets} (expressed with respect to comet 2P/Encke, denoted by subscript E)
\label{tab:Xparam}}
\tablecolumns{6}
\tablewidth{0pt}
\tablehead{
\colhead{\hspace{-20ex} comet name} &
\colhead{$| \Delta P|  /  |\Delta P_{\rm E}|$  }&
\colhead {$R/R_{\rm E}$ }&
\colhead{$P/P_{\rm E}$} &
\colhead{$\zeta /\zeta_{\rm E}$ }& 
\colhead{$X/X_{\rm E}$} 
}
\startdata
2P/Encke & 1 & 1 & 1 & 1 & 1 \\
9P/Tempel 1 & 3.50 & 1.18 & 3.73 & 0.35 & 1.0 \\
10P/Tempel 2 & 0.07 & 2.50 & 0.82 &  0.37 & 1.7 \\
19P/Borrelly &\hspace{-1.9ex} $>$10 & 1.04 & 2.55 & 0.35 & $\!\!\!\!>$4.8 \\
67P/Churyumov-Gerasimenko & 5.30 & 0.69 & 1.09 & 0.38 & 5.5 \\
103P/Hartley 2 &\hspace{-1.9ex} 37.50 & 0.24 & 1.64 & 0.43 & 1.9 \\ 
\enddata
\end{deluxetable*}

Table~\ref{tab:othercomets}
lists for a large sample of periodic comets the corresponding perihelion distance $q$, eccentricity $e$, rotation period $P$, effective radius $R$, $\zeta / \zeta_{\rm E}$, and the calculated $|\Delta P|$ per orbit for $X/X_{\rm E}$\,$=$\,1, $X/X_{\rm E}$\,$=$\,2.4, and $X/X_{\rm E}$\,$=$\,5.5 spanning the range of the parameter $X/X_{\rm E}$ in Table~\ref{tab:Xparam}. $X/X_{\rm E}$\,$=$\,2.4 is the geometric mean while excluding $X/X_{\rm E}$ for comet 2P/Encke (as $X$ for each comet was calculated with respect to that of 2P/Encke). The comets and physical parameters were selected primarily from table 1 of \citet{Rosita17} and references therein. However, only objects with quoted rotation periods and radii with reasonable error bars were used in our Table~\ref{tab:othercomets}. We do not quote the error bars in Table~\ref{tab:othercomets}, as they are small enough (a few tens of percent or less) for $P$ and $R$ and negligible for $q$ and  $\zeta / \zeta_{\rm E}$. (We refer the reader to \citet{Rosita17} and references therein for the measurement errors.)  Not all the objects we consider in Table~\ref{tab:othercomets} are  from table~1 of \citet{Rosita17}. Some are from our  Tables~\ref{tab:basecomets} and \ref{tab:Xparam} and those values are not based on those of \citet{Rosita17}.

Among the objects in table~1 of \citet{Rosita17}, comet 31P/Schwassmann-Wachmann~2 was excluded from Table~\ref{tab:othercomets} as its perihelion distance is large (3.4\,au). Therefore, this will  not result in a reasonable $\zeta$ value as $\zeta$ is calculated out to only 4\,au (i.e.\ the water production beyond 4\,au is ignored).  46P/Wirtanen is excluded as its rotation period is not well determined and is likely to be in a non-principal axis spin state \citep{Sam96}.   73P/Schwassmann-Wachmann~3-C was excluded as incompatible rotation periods are quoted in the literature \citep{Toth05, Storm06, Drahus10, Dykhuis12}. Objects not listed in  \citet{Rosita17} but included in our Table~\ref{tab:othercomets} are the Halley-type comets 1P/Halley,  96P/Machholz~1\footnote{Its orbital period is 5.3\,years and is uncharacteristic of Halley-type comets.}, and 109P/Swift-Tuttle, and the Jupiter family comet 45P/Honda-Mrkos-Pajdu\v{s}\'{a}kov\'{a}. Values for 1P/Halley are from \citet{Sam04} for the rotation period and from \citet{Belton91} for the effective radius. 1P/Halley is in a non-principal axis rotation state with component periods of 3.69\,day and 7.1\,day. We used an effective rotation period of 2.84\,day which is based on the vectorial sum of the two angular velocity components.  Values for comets 96P/Machholz~1 and 109P/Swift-Tuttle are from \citet{Meech96}, and references therein. Values for 45P/Honda-Mrkos-Pajdu\v{s}\'{a}kov\'{a} are from Arecibo radar observations by \citet{LH17}.

\begin{deluxetable*}{lcccccccc}[pbth]
\tablecaption{Heliocentric distances $q$, eccentricities $e$, rotation periods $P$, nuclear radii $R$, waster production ratios $\zeta / \zeta_{\rm E}$, and calculated changes in period $\Delta P$ for several values of $x$=$X/X_{\rm E}$ for 24 periodic comets. \label{tab:othercomets}}
\tablecolumns{9}
\tablewidth{0pt}
\tablehead{
\colhead{\hspace{-20ex} comet name} &
\colhead{$q$} &
\colhead{$e$} &
\colhead{$P$} &
\colhead {$R$ }&
\colhead{$\zeta /\zeta_{\rm E}$ }& 
\colhead{$| \Delta P(x$=1.0)$|$} &
\colhead{$| \Delta P(x$=2.4)$|$} &
\colhead{$| \Delta P(x$=5.5)$|$} \\
\colhead{} &
\colhead{[au]} &
\colhead{} &
\colhead{[hours]} &
\colhead{[km]} &
\colhead{} &
\colhead{[min]} &
\colhead{[min]} &
\colhead{[min]} 
}
\startdata
1P/Halley & 0.57 & 0.97 & 68.2 & 5.3 & 0.58 &  \hspace{-2ex} 18.2 &  \hspace{-2ex} 43.6 & 99.9 \\
6P/d'Arrest & 1.36 & 0.61 & \hspace{0.4ex} 6.7 & 2.2 & 0.35 & 0.6 & 1.5 & \hspace{0.4ex} 3.4 \\
7P/Pons-Winnecke &1.26 & 0.63 & \hspace{0.4ex} 7.9 & 2.6 & 0.38 & 0.7 & 1.6 &  \hspace{0.4ex} 3.7 \\
14P/Wolf & 2.41 & 0.41 & \hspace{0.4ex} 9.0 & 3.0 & 0.14 & 0.2 & 0.6 & \hspace{0.4ex} 1.3 \\
21P/Giacobini-Zinner & 1.03 & 0.71 & \hspace{0.4ex}  9.5 & 1.0 & 0.43 & 7.5 &  \hspace{-2ex} 17.9 & 41.1 \\
22P/Kopff & 1.58 & 0.54 & 12.3 & 2.2 & 0.31 & 1.9 & 4.4 & 10.2 \\
28P/Neujmin 1 & 1.55 & 0.78 & 12.8 & \hspace{-2ex} 10.7 & 0.25 & 0.1 & 0.2 & \hspace{0.4ex} 0.4 \\
45P/Honda-Mrkos-Pajdu\v{s}\'{a}kov\'{a} & 0.53 & 0.82 &  \hspace{0.4ex} 7.6 & 0.7 & 0.71 &  \hspace{-2ex} 18.6 &  \hspace{-2ex} 44.6 & \hspace{-1.6ex} 102.3 \\
47P/Ashbrook-Jackson & 2.31 & 0.40 & 15.6 & 3.1 & 0.14 & 0.7 & 1.6 & \hspace{0.4ex} 3.6 \\
48P/Johnson & 2.30 & 0.37 & 29.0 & 3.0 & 0.21 & 3.7 & 8.8 & 20.2 \\
49P/Arend-Rigaux & 1.42 & 0.60 &13.5 & 4.2 & 0.34 & 0.7 & 1.6 & \hspace{0.4ex} 3.6 \\
61P/Shajn-Schaldach & 2.11 & 0.43 & \hspace{0.4ex} 4.9 & 0.6 & 0.23 & 2.9 & 6.9 & 15.8 \\
76P/West-Kohoutek-Ikemura & 1.60 & 0.54 & \hspace{0.4ex}  6.6 & 0.3 & 0.31 &  \hspace{-2ex} 28.4 &  \hspace{-2ex} 68.0 & \hspace{-1.6ex} 155.9 \\
81P/Wild 2 & 1.60 & 0.54 &13.5 & 2.0 & 0.31 & 2.7 & 6.4 & 14.7 \\
92P/Sanguin & 1.83 & 0.66 & \hspace{0.4ex}  6.2 & 2.1 & 0.22 & 0.4 & 0.9 & \hspace{0.4ex} 2.0 \\
94P/Russell 4 & 2.24 & 0.36 & 20.7 & 2.3 & 0.22 & 3.4 & 8.2 & 18.9 \\
96P/Machholz 1 &0.12 & 0.96 &\hspace{0.4ex}  6.4 & 2.8 & 1.67 & 1.7 & 4.0 &  \hspace{0.4ex} 9.1 \\
109P/Swift-Tuttle & 0.97 & 0.96 & 69.4 &  \hspace{-2ex} 11.8 & 0.37 & 2.5 & 5.9 & 13.6 \\
110P/Hartley 3 & 2.48 & 0.31 & 10.2 & 2.3 & 0.20 & 0.7 & 1.8 &  \hspace{0.4ex} 4.0 \\
143P/Kowal-Mrkos & 2.54 & 0.41 &17.2 & 4.8 & 0.15 & 0.4 & 0.9 &  \hspace{0.4ex} 2.1 \\
162P/Siding Spring & 1.23 & 0.60 & 32.9 & 7.0 & 0.41 & 1.7 & 4.1 &  \hspace{0.4ex} 9.4 \\
169P/NEAT & 0.61 & 0.77 & \hspace{0.4ex} 8.4 & 2.5 & 0.69 & 1.5 & 3.5 &  \hspace{0.4ex} 8.1 \\
209P/LINEAR &0.97 & 0.67 & 10.9 & 1.5 & 0.51 & 5.2 &  \hspace{-2ex} 12.4 & 28.4 \\
260P/McNaught & 1.47 & 0.60 & \hspace{0.4ex}  8.2 & 1.5 & 0.31 & 1.8 & 4.3 &  \hspace{0.4ex} 9.8 \\
\enddata
\end{deluxetable*}

\begin{figure}[htbp]
\includegraphics[height=\textwidth,angle=270]{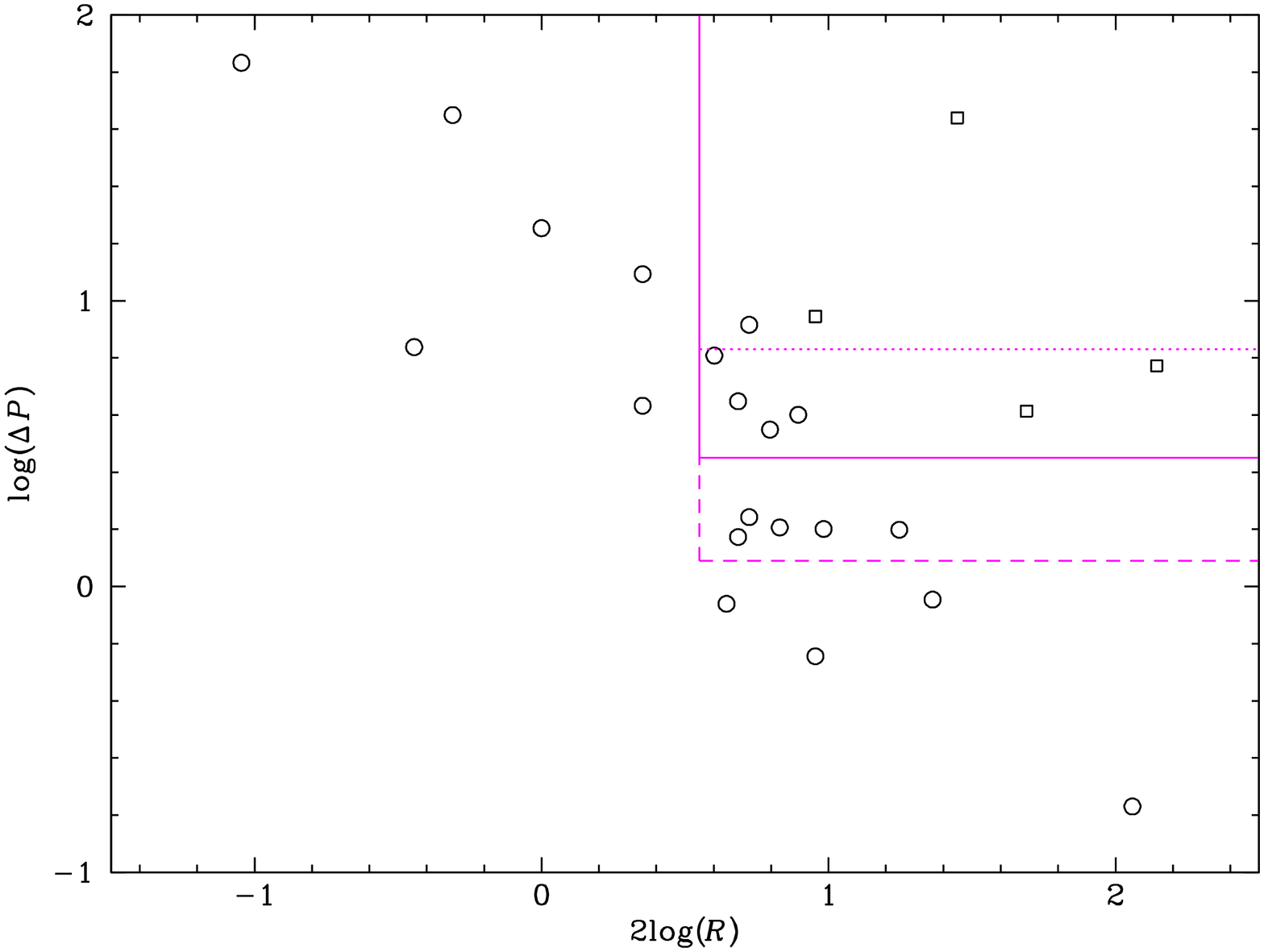}
\caption{Plot of $\log (\Delta P)$ versus $2 \log (R)$, using the values for $\Delta P$ calculated in Table~\ref{tab:othercomets} for $X/X_{\rm E}$\,$=$\,2.4. The comets occupying the upper right corner represent the brightest objects (assuming the same observing geometry and albedo) having the largest change in period $\Delta P$. The comets with rotation periods $>$1\,day are
shown with squares. Objects with long periods such as those require long 
observing runs to determine the respective rotation periods. However, they do 
show larger changes in rotation periods than fast rotators. The box with solid lines in the upper right encloses cometary candidates with the best chance of observing a period change for $X/X_{\rm E}$\,$=$\,2.4 The short dashed line denotes the lower boundary for $X$\,$=$\,1, and the long-dashed lines for $X/X_{\rm E}$\,$=$\,5.5. $\Delta P$ is expressed in minutes and $R$ in km. \label{fig:deltap}}
\end{figure}

We plot the predicted change in period ($\Delta P$) for $X/X_{\rm E}$\,$=$\,2.4 versus the square of the radius ($R^2$) as a log-log plot in Figure~\ref{fig:deltap}.  Assuming a similar observing geometry and albedo, as $R^2$ is a proxy for the flux,  we can then identify the candidates which will be the best to observe in order to  detect a period change. 
We expect that it is easier to observe rotation periods and measure rotational changes for comets having larger $\Delta P$ and larger nuclei, and they are therefore  the best candidates. 
Notice, that the  predicted changes in rotation periods  in Figure~\ref{fig:deltap} span more than  2.5 orders of magnitude (i.e., nearly a factor 300). Objects with smaller radii generally have larger changes in period but they are  faint and not easily observed.

We omit the measurement uncertainties in our figures to avoid cluttering the plots. For the selected objects, the uncertainties in the rotation periods and the radii are typically of the order of a few tens of percent or less while the ranges for rotation periods and radii in the plots span nearly 1.5 orders of magnitude (i.e., more than a factor of 30). 

For $X/X_{\rm E}$\,$=$\,1, the best candidates  to observe are 1P/Halley, 48P/Johnson, and 94P/Russell~4. For $X/X_{\rm E}$\,$=$\,2.4, the best candidates in addition to the ones mentioned for $X/X_{\rm E}$\,$=$\,1 are 22P/Kopff, 81P/Wild~2, 96P/Machholz~1, 109P/Swift-Tuttle, 162P/Siding~Spring, and 169P/NEAT. For $X/X_{\rm E}$\,$=$\,5.5 additional candidates are 6P/d'Arrest, 7P/Pons-Winnecke, 47P/Ashbrook-Jackson, 49P/Arend-Rigaux, and 110P/Hartley~3. We encourage the community to improve the accuracy of the rotation periods and measure the rotational changes for these comets.

1P/Halley was extensively observed from the ground during the 1986 apparition and no change in its period was noted. However, due to the long component periods and the excited spin, a change in rotation period of less than about an hour would have been difficult to observe. This would indicate that $X/X_{\rm E}$ is probably smaller than about 3. Recently, \citet{Eisner17} observed comet 49P/Arend-Rigaux in the 2011/2012 apparition and analyzed publicly available data from the 1984/1985 apparition and found that a possible change in rotation period has to be less than 14\,sec per orbit. This would point to  $X/X_{\rm E}$\,$<$\,1  for 49P/Arend-Rigaux. 

Comet 46P/Wirtanen will have a close approach to Earth (0.08\,AU) on UT  December 16, 2018. It is a very active comet \citep[e.g.][]{A'Hearn95} with a small nucleus having a radius  of $\approx$0.6\,km \citep[e.g.][]{Lamy04} and is very likely in a non-principal axis rotational state \citep{Sam96}. There are estimates of its rotation period from previous observations consistent with 6\,hours to 7.6\,hours \citep{Meech97, Lamy98, Boehn02}. These data were sparse or did not show convincing rotational phase plots. However, we can use  this range of periods and estimate the expected change in its rotation period. The full range of expected changes in rotation period for comet 46P/Wirtanen is 9\,min to 76\,min with a most likely change of about half-an-hour. However, as the spin state of 46P/Wirtanen is not well determined, the rotation period or component periods could be larger (it is unlikely that the period is much smaller than 6\,hours) which would increase the expected change in its rotation period even more. If 46P/Wirtanen behaves similarly to comet 103P/Hartley~2, such a change would be easily observable in the coming apparition with coma morphological studies.

\section{Correlations among orbital and physical parameters} \label{sec:corr}

We can also explore correlations among other orbital and physical parameters. Figure~\ref{fig:zetaq} shows $\log (\zeta)$ versus $\log (q^{-1})$ for the objects from Tables~\ref{tab:basecomets} and \ref{tab:othercomets}.  As mentioned previously, error bars are omitted in the figures. The parameter $\zeta$ is based on the theoretical water production rate as a function of heliocentric distance derived from the thermo-physical model of \citet{Pedro01}, and as water is the dominant sublimating volatile at the heliocentric distances considered, the uncertainties in $\zeta$ are systematic and of the order of a few percent at most. For comparison, the range of $\zeta$ for the different comets we consider is an order of magnitude. The error bars for the perihelion distance $q$ are  negligible since the orbits are well-determined and  they would be too small to be shown. There is a linear correlation between  $\log (\zeta)$ and $\log (q^{-1})$ with a slope of 0.81$\pm$0.05 and an intercept of 2.83$\pm$0.01. Therefore, to first order, $q^{-0.8}$ with respect to that of 2P/Encke could be substituted for $\zeta/\zeta_{\rm E}$. This is a time saver  for many calculations as the perihelion distance is readily available while $\zeta$ has to be numerically determined. It is notable that to first order, the total water production per unit surface area per orbit is correlated with the perihelion distance, regardless of the eccentricity. We note from Figure~\ref{fig:zetaq} that on average objects with smaller eccentricity have lower water production, which is really not a surprising result as they typically have larger perihelion distances. 

A plot similar to Figure \ref{fig:zetaq} could be generated for all known periodic comets. However, as our sample is essentially random (since we are using objects with known rotation periods but without any consideration for their orbital parameters), and since the range of perihelia covers the entire range for periodic comets, we do not expect such a plot to yield different results. 

In addition, this relationship between $\zeta$ and $q$ will provide an alternate way to estimate the orbitally averaged effective active fraction ($f$) of a comet. $f$ can be estimated from
\begin{equation} \label{eq:activef}
f = \frac{M_W}{10^{2.83} 4 \pi R^2 q^{-0.8}} \approx \frac{M_W}{8500 R^2 q^{-0.8}}
\end{equation}
 where $M_W$  (in grams) is the {\it observed} total water production from the nucleus per orbit, $R$ (in cm) is the radius of the nucleus, and $q$ (in au) is the perihelion distance of the comet. We can use comet 67P/Churyumov-Gerasimenko as an example. Using the values of the radius and perihelion distance from our Table~\ref{tab:basecomets} and $M_W$\,=\,6.4$\times$10$^{12}$\,grams \citep{Hansen16}, we derive $f$\,=\,0.032. \citet{Hansen16} also provide a value for the shape-model based surface area of 46.6\,km$^2$ which yields $f$\,=\,0.024. These values compare well with active fractions of 0.014 derived by \citet{Snodgrass13} and of $<$0.07 cited by \citet{Lamy07}.

 Figure~\ref{fig:Pzeta} shows the period $P$ versus $\zeta$ as a log-log plot with different symbols and colors for various radius ($R$) bins. Error bars are again omitted. This figure shows that there is no clear correlation between the rotation period and the
amount of water production per unit surface area per orbit. However, it can be seen that the largest rotation periods (i.e., slow rotators) seem to correlate with large nuclei. In order to further explore this,  we show the rotation period $P$ versus nuclear radius $R$
as log-log and linear plots in Figure~\ref{fig:PR}.  There are many objects in common in our Figure~\ref{fig:PR} and figure 61 from \citet{Rosita17}, however, we did not consider objects with large error bars.  \citet{Rosita17} discuss the lack of objects in the lower right portion of their plot but do not further explore the implications. It is clear that there is a definite correlation between the rotation period
and the nuclear radius despite the scatter. Again, on average, larger objects seem
to have larger rotation periods. This is not surprising as it is harder to change
the rotation period of larger (i.e.\ more massive) objects, whereas smaller objects undergo rapid rotational 
changes. Such rapid rotational changes will effectively result in rotational spin-ups \citep[e.g.][]{Whipple82, SB95}, thus causing small nuclei to end up showing rapid rotational rates (i.e., small rotation periods). Alternatively, as fast rotators take a long time to change their rotational rates, on average most small nuclei show  comparatively  smaller rotation periods than large nuclei.

\section{Discussion}\label{sec:disc}

Adding two more Jupiter family comets to the four previously used to derive the parameter $X$ confirms that it is  nearly constant within a factor of a few despite the increased scatter. This implies that we can use this formalism to identify comets that are expected to have observable changes in their rotation periods. It is more difficult to obtain good rotational coverage from the ground for comets with very long rotation periods ($>$1\,day) and they might therefore not be the best candidates for determining changes in the rotation periods. Such objects are identified in Figure~\ref{fig:deltap} with square symbols.
Even though long rotation periods require long observing runs, the long rotation periods themselves make it easier to change the respective rotation periods. The timescale $\tau$ for the change in rotation period due to outgassing torques  is 
\begin{equation} \label{eq:tau}
\tau \propto \frac{R^4} {\dot{m}P}
\end{equation}
 \citep[e.g.][]{Sam86,Jewitt97} where $\dot{m}$ is the mass loss rate due to outgassing from all active regions on the nuclear surface. Therefore, 
  \begin{equation}\label{eq:var}
 \tau  \propto \frac{R^{4-n}}{P}
 \end{equation}
with 0\,$\le$\,$n$\,$\le$\,2.
If the outgassing is from the entire surface area or a fixed active fraction, then $\dot{m}$\,$\propto$\,$R^2$ (i.e.\ $n$\,$=$\,2). On the other hand,  although less likely, if the outgassing area is the same irrespective of the nuclear size, then $\dot{m}$\,$\propto$\,$R^0$ (i.e.\ $n$\,$=$\,0). An interesting result from this investigation is the ensemble behavior between the rotation periods and the radii of the nuclei. As expected from the functional dependency of the timescale for the sublimation-driven rotational change in the period (equation~\ref{eq:var}), small nuclei would on average undergo rapid changes to their rotation periods. As a result of this rotational evolution,  many of them ultimately spin up \citep{Whipple82, Sam04}. This is more apparent for small nuclei than larger ones as can be easily seen in the left panel of Figure~\ref{fig:PR}.  Furthermore, we note that the range in rotation periods tends to be smaller for small nuclei than large nuclei  (right panel of Figure~\ref{fig:PR}). This sublimation-driven spin-up is likely to be a major contributor to the ultimate break-up of many cometary nuclei, especially for the smaller objects.

The linear relationship shown in Figure~\ref{fig:zetaq} between $\log(\zeta)$ and $\log(q^{-1})$ with little scatter was a surprising result. However, we anticipated a general trend of increasing $\zeta$ with increasing $q^{-1}$ but with a large vertical scatter as there was a distribution of eccentricities  for the orbits of the  comets considered. The fact that there is a linear correlation between $\log{\zeta}$  and  $\log(q^{-1})$ means that we can adopt the perihelion distance $q$ (or more specifically $q^{-0.8}$) as a proxy for the total water production of a comet per unit surface area. The maximum deviation of $\zeta $ from the fit is less than about 50\% (i.e.\ the maximum vertical offset to the fit in the $\log$ plot is less than 0.18), and the average scatter is much smaller. We expect the scatter to be due to differences in orbital characteristics caused by orbital parameters such as eccentricities. We provide a rationalization of this relationship between $\zeta$ and $q$ in  Appendix~A.

\section{Conclusions} \label{sec:conc}

\begin{itemize}
\item We expanded the number of comets with confirmed changes in rotation periods from 4 to 6 for estimating the parameter $X$. It is still constant within a factor of a few  albeit with a larger scatter.

\item We determined the expected changes in rotation periods for 24 periodic comets using a range of $X$ and identified the best candidates to observe in order to determine additional values for rotation changes.

\item We found a linear correlation between $\log(\zeta)$ and $\log(q^{-1})$ with a slope of 0.8, indicating $\zeta$\,$\propto$\,$q^{-0.8}$. A theoretical argument rationalizing this is provided in Appendix~A. We further note that comets with low eccentricities typically  have higher perihelion distances.

\item We found that objects with  smaller radii are more likely to be observed with short rotation periods, consistent with what is expected for sublimation-driven rotational changes.

\end{itemize}

\acknowledgements
We gratefully acknowledge support in part from NASA Planetary Astronomy Grant \#NNX14AG73G, from NASA Discovery Data Analysis Grant \#NNX15AL66G, and from NASA Solar System Observations Grant \#NNX16AG70G.  We thank the referee for a thorough review with suggestions that improved this manuscript. We thank Dr.\ Pedro Guti\'{e}rrez for providing the water production rates as a function of heliocentric distance based on his thermo-physical model in digital form.

\appendix

\section{Rationalization of the relationship between $\zeta$ and the perihelia}

Here we provide a rationalization of the functional relationship between $\zeta$ and $q$. specifically,  $\zeta$\,$\propto$\,$q^{-0.8}$. A direct analytical derivation of the exact functional dependence of $\zeta$ on $q$ is complicated since the water production rate has a complex functional dependence on the heliocentric distance $r_{\rm h}$. For example, the dimensionless empirical water sublimation rate $g(r_{\rm h})$ suggested by \citet{Marsden73} based on the input from Delsemme \& Delsemme (1971; personal communication) has the following functional form:
\begin{equation}
g(r_{\rm h}) = 0.1113 \bigg(\frac{r_{\rm h}}{2.808}\bigg)^{-2.15} \bigg[1+\bigg(\frac{r_{\rm h}}{2.808}\bigg)^{5.093}\bigg]^{-4.6142}
\end{equation}
As can be seen from this equation, when close to the sun (i.e., $r_{\rm h}$\,$\ll$\,2.808\,au), the term inside the square bracket approaches 1 and $g(r_{\rm h})$ is nearly proportional to $r_{\rm h}^{-2.15}$, whereas when $r_{\rm h}$\,$\gg$\,2.808\,au, $g(r_{\rm h})$ falls off extremely rapidly. This is simply because, as a comet is far away, a significant fraction of the incident solar energy is spent for thermal re-radiation and heating up the nucleus. On the other hand, when the comet is close to the sun, a larger fraction of the incident energy is spent on sublimation of water ice. The amount of energy corresponding to reflection is a constant fraction regardless of the heliocentric distance and depends on the albedo, which typically is of the order of 4\% for dark objects such as comets.

\begin{figure}[htbp]
\includegraphics[height=\textwidth,angle=270]{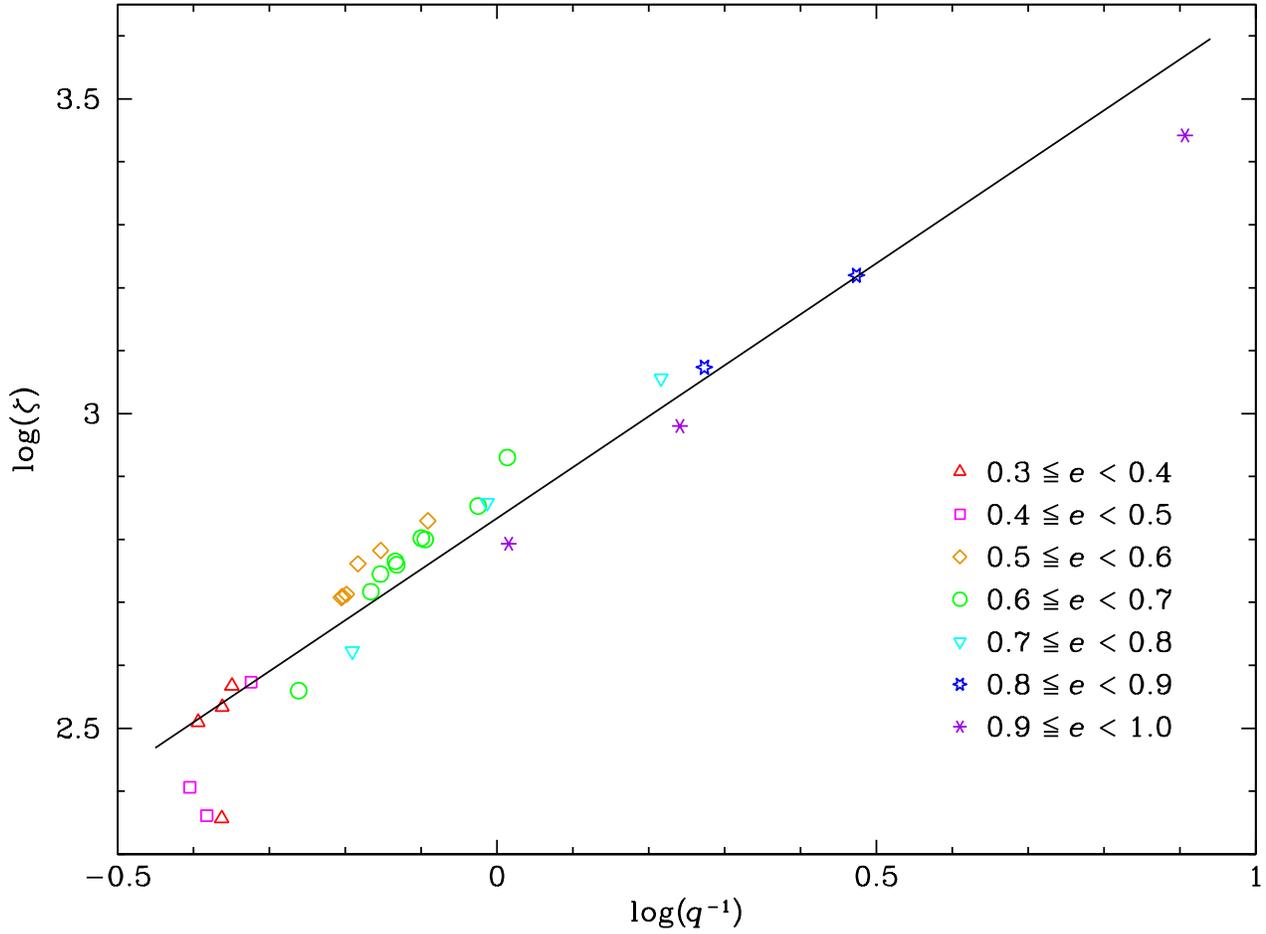}
\caption{Plot of $\log (\zeta)$ versus $\log (q^{-1})$ for the objects from Tables~\ref{tab:basecomets} and \ref{tab:othercomets} with a linear fit (black line) with a slope of 0.81 and an intercept of 2.83. On average, objects with smaller eccentricities tend to have lower total water production per unit surface area per orbit. The objects with eccentricities larger than 0.9 (purple asterisks)  in this plot are all Halley-type comets. 
There is no clear difference in the $\zeta$ versus $q^{-1}$ behavior between the Halley-type comets and the Jupiter family comets listed in our Table~\ref{tab:othercomets}.   $\zeta$ is the total water production per unit surface area over the active phase of the orbit. The perihelia are given in au and $\zeta$ in g\,cm$^{-2}$. \label{fig:zetaq}}
\end{figure}

\begin{figure}[htbp]
\includegraphics[height=\textwidth,angle=270]{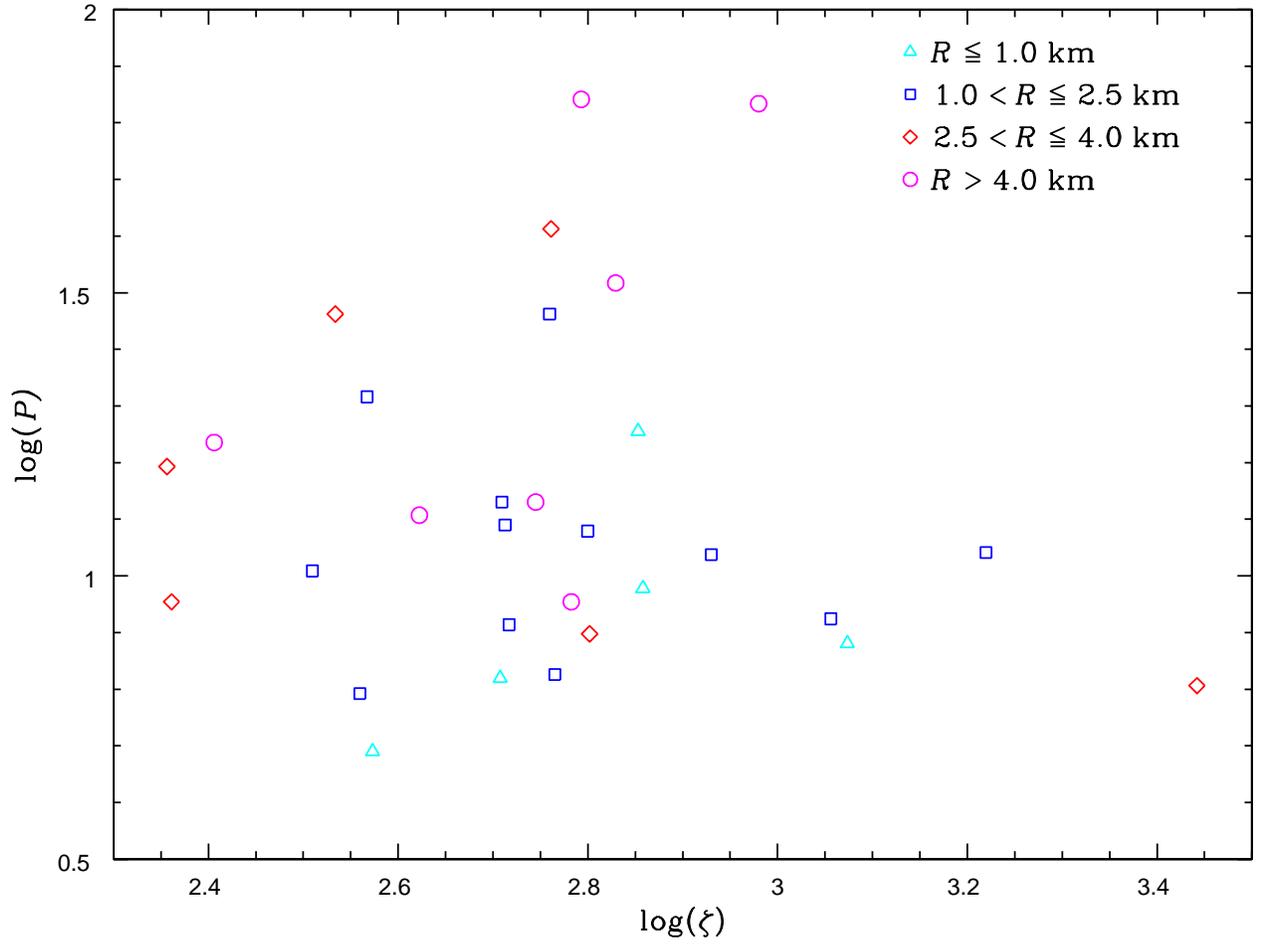}
\caption{Plot of  $\log (P)$ versus $\log (\zeta)$ for the objects from Tables~\ref{tab:basecomets} and \ref{tab:othercomets} with different symbols and colors for various radius ($R$) bins with the legend  given inside the figure.  The periods are expressed in hours and $\zeta$ in g\,cm$^{-2}$. \label{fig:Pzeta}}
\end{figure}

\begin{figure}[htbp]
\includegraphics[height=0.5\textwidth,angle=270]{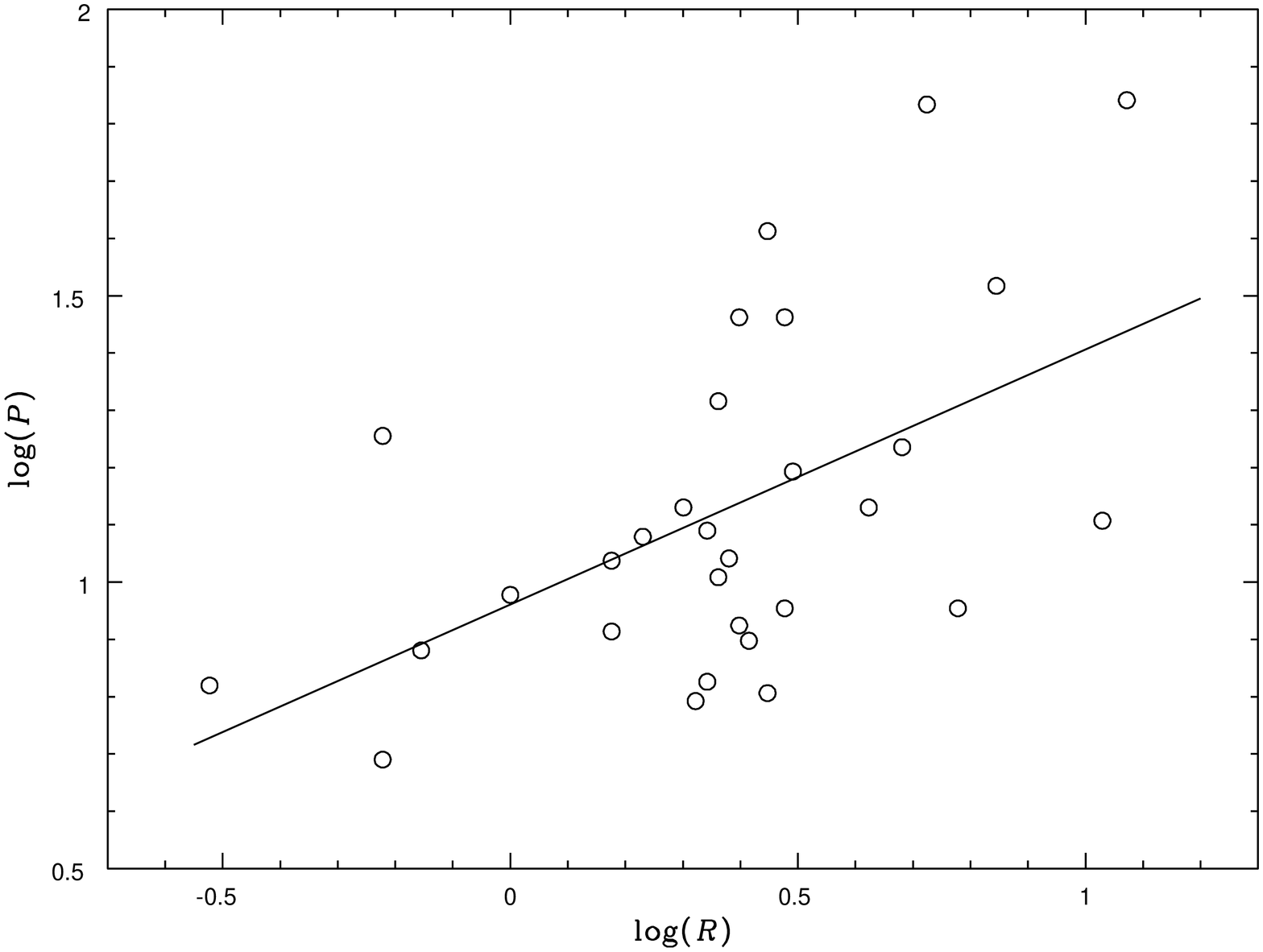}\includegraphics[height=0.5\textwidth,angle=270]{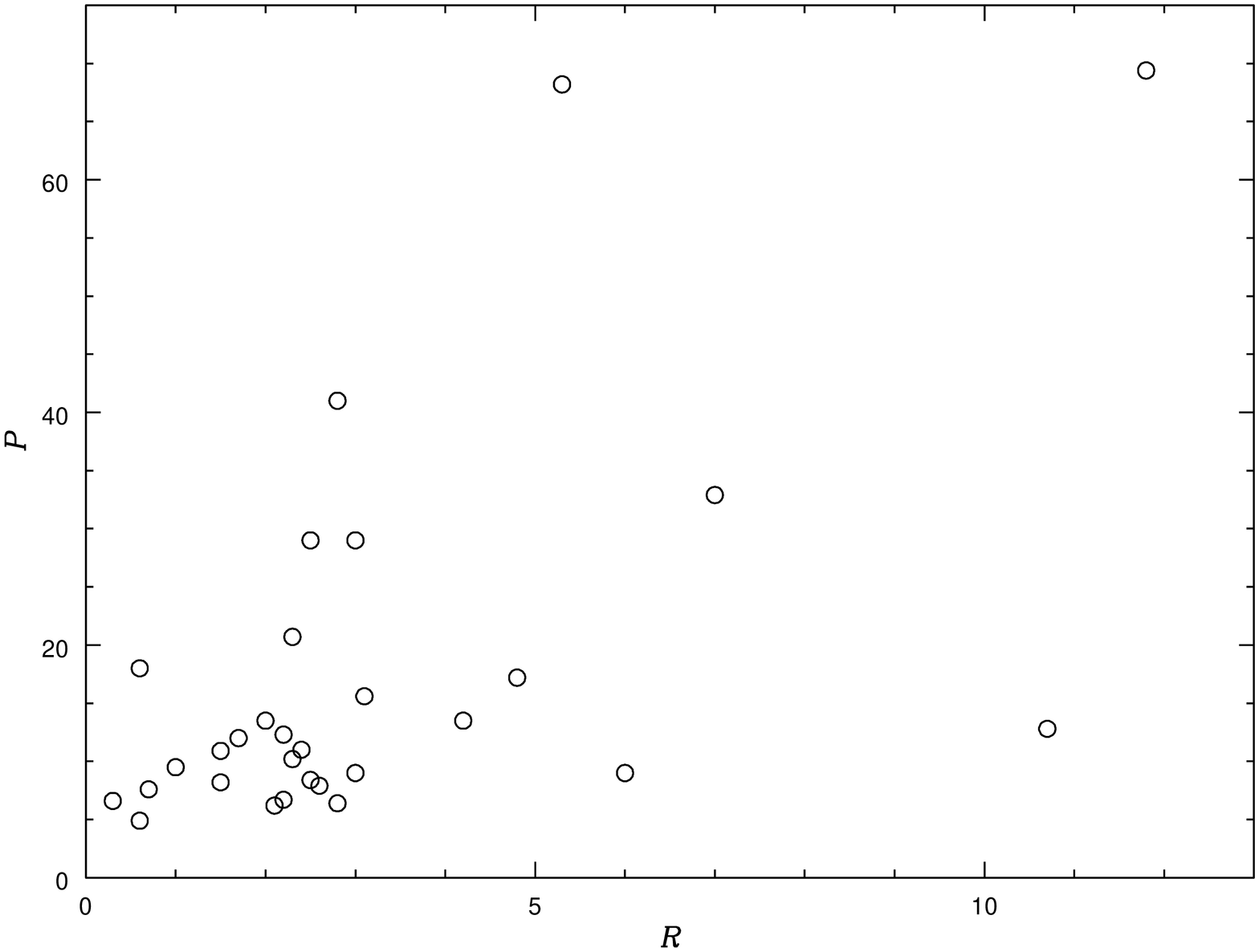}
\caption{ {\it Left:} Plot of $\log (P)$ versus $\log (R)$ for the objects from Tables~\ref{tab:basecomets} and \ref{tab:othercomets} with a linear fit (black line) with a slope of 0.56. The fit is just meant to guide the eye and we do not imply that the correlation is linear.  {\it Right:} The same plot as on the left is shown in linear dimensions, so that it can be more easily discerned that the range in periods tends to be small for smaller radii. The rotation periods are given in hours and the radii in km.  \label{fig:PR}}
\end{figure}

We look at the average energy incident on the nucleus per unit surface area per unit time over an orbit. This quantity is proportional to the reciprocal of the products of the semi-major and semi-minor axes of the orbit \citep[e.g.][p.137]{Danby88}. That is,
\begin{equation}\label{eq:Danby}
\frac{E}{P_{\rm orb}} \propto \frac{1}{ab}
\end{equation}
where $E$ is the energy incident on the nucleus per unit surface area over an orbit, $P_{\rm orb}$ is the orbital period, $a$ is the semi-major axis, and $b$ is the semi-minor axis.
From Kepler's third law
\begin{equation}\label{eq:Kepler}
P_{\rm orb} \propto a^{3/2}
\end{equation}
and from the properties of an ellipse
\begin{equation}\label{eq:ellipse}
b = a(1-e^2)^{1/2}
\end{equation}
Substituting equations \ref{eq:Kepler} and \ref{eq:ellipse} into \ref{eq:Danby}, yields
\begin{equation}
E \propto  \frac{a^{-1/2}}{(1-e^2)^{1/2}}
\end{equation}
As the perihelion distance $q$ and the semi-major axis $a$ are related by
 \begin{equation}  
 a = \frac{q}{1-e}  
 \end{equation}     
 we obtain
 \begin{equation}    
 E \propto  q^{-1/2}                                                                                                         
 \end{equation}                                                                                                                    
If the entire incident energy (or a constant fraction of it after accounting for the reflected energy) is spent on water sublimation, irrespective of the heliocentric distance, then we should expect 
 \begin{equation}   
 \zeta \propto q^{-0.5}                                                                                                           
 \end{equation} 
However, when a comet moves further away from the sun, progressively smaller fractions of the incident energy are spent on water sublimation. Therefore, it is natural for $\zeta$ to show a more rapid fall-off than $q^{-0.5}$ thus rationalizing the $q^{-0.8}$ behavior we derived from Figure~\ref{fig:zetaq}.

\end{document}